\documentclass[12pt]{article}
\usepackage{curves}
\newcommand{\dd}{\mbox{\rm d}}

\newcommand{\oo}{\over}
\newcommand{\p}{\partial}

\newcommand{\be}{\begin{equation}}
\newcommand{\ee}{\end{equation}}
\newcommand{\lbl}{\label}
\newcommand{\bi}{\bibitem}

\newcommand{\vs}{\vspace}

\newcommand{\ci}{\cite}

\def\bear{\begin{eqnarray}}
\def\ear{\end{eqnarray}}

\begin{document}
\baselineskip .7cm

\thispagestyle{empty}

\begin{center}
{\bf \LARGE Clifford Algebra of Spacetime and the Conformal Group}
\end{center}

\vs{5mm}

\begin{center}

C. Castro$^a$\footnotetext{$^a$ Center for Theoretical Studies of Physical Systems,
Clark Atlanta University, Atlanta}
and M. Pav\v si\v c$^b$\footnotetext{
$^b$Jo\v zef Stefan Institute, Jamova 39, SI-1000 Ljubljana, Slovenia;
Email: matej.pavsic@ijs.si}

\end{center}

\vs{.5cm}

\centerline{ABSTRACT}

\vs{8mm}

We demonstrate the emergence of the conformal group SO(4,2) from the
Clifford algebra of spacetime. The latter algebra is a manifold, called
Clifford space, which is assumed to be the arena in which physics takes 
place.
A Clifford space does not contain only points (events), but also lines,
surfaces, volumes, etc..., and thus provides a framework for description of
extended objects. A subspace of the Clifford space is the space whose metric
is invariant with respect to
the conformal group SO(4,2)  which can  be given either passive or active
interpretation. As advocated long ago by one of us, active conformal
transformations, including dilatations, imply that sizes of physical
objects can change as a result of free motion, without the presence of 
forces.
This theory is conceptually and technically very different
from Weyl's theory and provides when extended to a curved conformal space
a resolution of the long standing problem of realistic masses in 
Kaluza-Klein
theories.

\vs{1cm}


\newpage

\section{Introduction}

Extended objects such as membranes or branes of any dimension are
nowadays subject of extensive studies. A deeper geometric principle behind
such theories remains to be fully explored. It has been found that
various string theories are different faces of a conjectured single theory,
called $M$-theory which has not yet been rigorously formulated. In a number
of recent works branes have been considered from the point of view of
the geometric calculus based on Clifford algebra
\ci{Castro}--\ci{Pavsic}. The latter is a very
useful tool for description of geometry and physics
\ci{Castro}--\ci{Castro-Pavsic}. A space (or spacetime)
consists of points (or events). But besides points there are also lines,
surfaces, volumes, etc.\ . Description of such geometric objects turns out
to be very elegant if one employs multivectors which are the outer products
of vectors. All those objects are elements of Clifford algebra. Since in
physics we do not consider point particles only, but also extended objects,
it appears natural to consider Clifford algebra as an arena in which physics
takes place. Clifford algebra as an arena arena for physics has been called
{\it pandimensional continuum} \cite{Pezzaglia} or $C$-{\it space} \cite
{Castro2}.

In this paper we report about a possibly far reaching observation that
Clifford algebra of 4-dimensional spacetime contains conformal group
SO(4,2) as a special case. It was proposed long time ago
\cite{Pavsic2,Pavsic3}
that conformal space can serve as the arena for physic which involves
active dilatations and dilatational motions,
and that a curved conformal
space is a possible realization for Kaluza-Klein theory. A remarkable
property of such an approach to Kaluza-Klein theories is that the notorious
problem of Planck mass does not occur, since the 4-dimensional
mass is given by the expression $m = \sqrt{m_{00} + \pi_5 \pi_6}$ in
which $m_{00}$ is the invariant mass in 6-dimensions,
$\pi_6$ the electric charge and $\pi_5$ the dilatational
momentum (taken to be zero for the ordinary electron). However in that
old work it was not yet realized that the variables $\kappa$ and $\lambda$
entering the description of SO(4,2) ---which obviously were not just
the ordinary extra dimensions--- are natural ingredients of the Clifford
algebra
of 4-dimensional spacetime: they are coordinates of $C$-space.

The idea to allow objects to change in size when they move in spacetime
is in fact very old and dates back to Herman Weyl \ci{Weyl}.
However, our proposal, initiated
in refs. \ci{Pavsic2,Pavsic3}, differs from that of Weyl both conceptually
and technically, and is free from the well-known Einstein's
criticism. In Weyl's geometry one is gauging the local scale transformations
by introducing a gauge field, identified with the electromagnetic field
potential, such that sizes of objects (and the rate at which clocks tick) 
are
path dependent. Einstein pointed out that this would effect spectral
lines emitted by the atoms which had followed different paths in spacetime 
and then brought together. A result would be blurred spectra with no distinctive
spectral lines, contrary to what we observe. In our approach we are not
gauging the local scale transformations and scale changes of objects are
not due to the different paths they traverse. Scale is postulated as
an extra degree of freedom, analogous to position. If the observer chooses 
to be in the same scale-frame of reference as the particular
object he observes (for more detailed
description see refs. \ci{Pavsic2,Pavsic3}), no scale changes for that object 
are observed (with respect to the observer) in its ordinary spacetime 
motion.
However, in general, our theory predicts that in a
given scale-frame of reference chosen by the observer, different
objects can have different scales (i.e., different sizes) and consequently
emit spectral lines whose wavelengths are shifted by the corresponding
scale factors. When considering the quantized theory, not only position
but also scale has to be quantized \ci{Pavsic2,Pavsic3} which affects not
only relative positions but also relatives scales of
bound objects, e.g., the atoms within a crystal. A crystal as a free object
can have arbitrary position and scale, whilst the atoms bound within
the crystal have the relative positions and scales as determined by
the solutions of the Schr\" odinger equation (generalized to C-space).
Possible astrophysical implications of scale as a degree of freedom were
discussed refs. \ci{Pavsic2,Pavsic3}.\

\section{On geometry, Clifford algebra and physics}

Since the appearance of the seminal books by D. Hestenes \cite{Hestenes}
we have a very useful language and tool for geometry and physics,
which is being recognized by an increasing number of researchers
\cite{Lounesto}--\cite{Castro-Pavsic}. Although Clifford algebra
is widely used and explored both in mathematics and physics, its full
power for formulation of new physical theories has been recognized only
relatively recently \ci{Pezzaglia,Castro2,Pavsic, Pavsic1}.
For the reasons of self consistency we will provide a brief introduction
into the geometric calculus based on Clifford algebra and point
out how Minkowski spacetime can be generalized to Clifford space.

The starting observation is that the basis vectors $e_{\mu}$ in an
$n$-dimensional space $V_n$ satisfy the Clifford algebra relations
\be
       e_{\mu} \cdot e_{\nu} \equiv \mbox{${1\oo 2}$} (e_{\mu} e_{\nu}
       + e_{\nu} e_{\mu}) = g_{\mu \nu}
\lbl{1}
\ee
where $g_{\mu \nu}$ is the metric of $V_n$. The dot denotes the {\it inner
product} of two vectors. It is the symmetric part of the non commutative
{\it Clifford} or {\it geometric product}
\be
       e_{\mu} e_{\nu} = e_{\mu} \cdot e_{\nu} + e_{\mu} \wedge e_{\nu}
\lbl{2}
\ee
whose antisymmetric part is the {\it wedge} or {\it outer product}
\be
        e_{\mu} \wedge e_{\nu}  \equiv  \mbox{${1\oo 2}$} (e_{\mu} e_{\nu}
       - e_{\nu} e_{\mu}) \equiv {1\oo {2!}} [e_{\mu}, e_{\mu}]
\lbl{2a}
\ee
Whilst the inner product is a scalar, the outer product is a bivector.
It denotes an {\it oriented area} of a 2-surface enclosed by a 1-loop.
The precise shape of the loop is not determined. In a similar manner we
can form higher {\it multivectors} or $r$-{\it vectors}
\be
       e_{\mu_1} \wedge e_{\mu_2} \wedge ... \wedge e_{\mu_r} \equiv
       {1\oo {r!}} [e_{\mu_1}, e_{\mu_2},...,e_{\mu_r}]
\lbl{3}
\ee
by antisymmetrizing the Clifford product of $r$ vectors. Such an object
is interpreted geometrically to denote an oriented $r$-area of an
$r$-surface enclosed by an $(r-1)$-loop. Multivectors are elements of
the {\it Clifford algebra} ${\cal C}_n$ of $V_n$. 
An element of ${\cal C}_n$ is called a {\it Clifford number}. 

In the geometric (Clifford) product (\ref{2}) a scalar and a bivector
occur in the sum. In general, a Clifford number is a superposition,
called a {\it Clifford aggregate} or {\it polyvector}:
\be
     A = a + {1\oo {2!}} a^\mu e_\mu + {1\oo {3!}} a^{\mu \nu} \, e_\mu
     \wedge e_\nu + ... + {1\oo {n!}} a^{\mu_1 ...\mu_n} e_{\mu_1} \wedge 
...
     \wedge e_{\mu_n}
\lbl{4}
\ee
In an $n$-dimensional space an $n$-vector is a multivector of the highest
possible degree; an $(n+1)$-vector is identically zero.

Considering now a flat 4-dimensional spacetime with basis vectors
$\gamma_\mu$
satisfying
\be
         \gamma_\mu \cdot \gamma_\nu = \eta_{\mu \nu}
\lbl{4a}
\ee
where $\eta_{\mu \nu}$ is the diagonal metric with signature $(+ - - -)$
eq.(\ref{4})  reads
\be
       D = d + d^\mu \gamma_\mu + {1\oo {2!}} \gamma_\mu \wedge \gamma_\nu
       + {1\oo {3!}} d^{\mu \nu \rho} \gamma_\mu \wedge \gamma_\nu \wedge
       \gamma_\rho + {1\oo {4!}} d^{\mu \nu \rho \sigma}
       \gamma_\mu \wedge \gamma_\nu \wedge \gamma_\rho \wedge \gamma_\sigma
\lbl{5}
\ee
where $d$, $d^{\mu}$, $d^{\mu \nu}$,..., are scalar coefficients. The
Clifford
algebra in Minkowski space $V_4$ is called the {\it Dirac algebra}.

Let us introduce the symbol $I$ for the unit element of 4-volume
(pseudoscalar)
\be
        I \equiv \gamma_0 \wedge \gamma_1 \wedge \gamma_2 \wedge \gamma_3
        = \gamma_0 \gamma_1 \gamma_2 \gamma_3 \; \; , \qquad I^2 = -1
\lbl{6}
\ee
Using the relations
\be
         \gamma_\mu \wedge \gamma_\nu \wedge \gamma_\rho \wedge 
\gamma_\sigma
         = I \epsilon_{\mu \nu \rho \sigma}
\lbl{7}
\ee
\be
         \gamma_\mu \wedge \gamma_\nu \wedge \gamma_\rho =
         I \epsilon_{\mu \nu \rho \sigma} \gamma^\sigma
\lbl{8}
\ee
where $\epsilon_{\mu \nu \rho \sigma}$ is the totally antisymmetric tensor,
and introducing the new coefficients
\be
         S \equiv d \; , \quad V^\mu \equiv d^\mu \; , \quad T^{\mu \nu}
\equiv
         \mbox{${1\oo 2}$} d^{\mu \nu} \nonumber
\ee
\be
         C_\sigma \equiv \mbox{${1\oo {3!}}$} d^{\mu \nu \rho}
         \epsilon_{\mu \nu \rho \sigma} \; , \quad P \equiv \mbox{${1\oo
{4!}}$}
         d^{\mu \nu \rho \sigma} \epsilon_{\mu \nu \rho \sigma}
\lbl{9}
\ee
we can rewrite $D$ of eq.(\ref{5}) as the sum of a scalar, vector, bivector,
pseudovector and pseudoscalar:
\be
       D = S + V^{\mu} \gamma_\mu + T^{\mu \nu} \gamma\wedge \gamma_\nu +
       C^\mu \, I \gamma_\mu + P \, I
\lbl{10}
\ee

So far physics in spacetime has been predominantly using only the vector
part of $D$. The full Clifford algebra or Dirac algebra have been used in
relativistic quantum theory, but not in the classical special or general
relativity, neither in the theory of strings and branes.

Assuming that fundamental physical object are not point particles but
extended
objects such as strings and branes of arbitrary dimension, it has been
proposed \ci{Pezzaglia,Castro2,Pavsic,Pavsic1} that {\it physical quantities
such
as positions and velocities
of those objects are polyvectors}. It was proposed to rewrite the known
fundamental string and brane actions by employing polyvector coordinates
\be
       X = {1\oo {r!}} \sum_{r=0}^n X^{\mu_1...\mu_r} \gamma_{\mu_1} \wedge
       ...\wedge \gamma_{\mu_r} \equiv X^A E_A
\lbl{11}
\ee
Here we use a compact notation in which $X^A \equiv X^{\mu_1...\mu_r} $ are
{\it real} coordinates, and $E_A \equiv  \gamma_{\mu_1} \wedge...\wedge 
\gamma_{\mu_r}$
basis vectors of the $2^n$-dimensional {\it Clifford algebra} of spacetime.
The latter algebra of spacetime positions and corresponding higher grade
objects, namely oriented $r$-areas, is a manifold which is more general than
spacetime. In the literature such a manifold has been named {\it
pandimensional continuum} \ci{Pezzaglia} or {\it Clifford space} or
$C$-{\it space} \ci{Castro2}.

The infinitesimal element of position polyvector (\ref{11}) is
\be
         \dd X = {1\oo {r!}} \sum_{r=0}^n \dd X^{\mu_1...\mu_r}
         \gamma_{\mu_1} \wedge
       ...\wedge \gamma_{\mu_r} \equiv \dd X^A E_A
\lbl{12}
\ee
We will now calculate its norm squared. Using the definition for the {\it 
scalar
product} of two polyvectors $A$ and $B$
\be
         A * B = \langle A B \rangle_0
\lbl{13}
\ee
where $\langle \; \rangle_0$ means the scalar part of the geometric product
$A B$, we obtain
\be
        |\dd X |^2 \equiv \dd X^{\dagger} * \dd X = \dd X^A \, \dd X^B
        G_{AB} = \dd X^A \, \dd X_A
\lbl{14}
\ee
Here
\be
          G_{AB} = E_A^{\dagger} * E_B
\lbl{15}
\ee
is the $C$-space metric and $A^{\dagger}$ the reverse\footnote{Reversion or,
alternatively, {\it hermitian conjugation}, is the operation which reverses
the order of all products of vectors in a decomposition of a polyvector
$A$.}
of a polyvector $A$.

For example, if the indices assume the values $A=\mu, \; B=\nu$, we have
\be
        G_{\mu \nu} = \langle e_\mu e_\mu \rangle_0 = e_\mu \cdot e_\nu =
        g_{\mu \nu}
\lbl{16}
\ee
If $A = [\mu \nu]$, $B=[\alpha \beta]$
\bear
         G_{[\mu \nu][\alpha \beta]} &=& \langle (e_\mu \wedge 
e_\nu)^\dagger
         (e_\alpha \wedge e_\beta \rangle_0 =
         \langle (e_\mu \wedge e_\nu)^\dagger \cdot (e_\alpha
         \wedge e_\beta \rangle_0 \nonumber \\
          &=& (e_\mu \cdot e_\alpha)(e_\nu \cdot e_\beta)
         - (e_\nu \cdot e_\alpha)(e_\mu \cdot e_\beta) = g_{\mu \alpha}
         g_{\nu \beta} - g_{\nu \alpha} g_{\mu \beta}
\lbl{17}
\ear
If $A=\mu$, $B= [\alpha \beta]$
\be
         G_{\mu [\alpha \beta]} = \langle e_\mu (e_\alpha \wedge e_\beta)
         \rangle_0 = 0
\lbl{18}
\ee

Explicitly we have
\bear
        |\dd X |^2 &=& {1\oo {r!}} \sum_{r=0}^n \dd  X^{\mu_1...\mu_r}
        \dd  X_{\mu_1...\mu_r} \nonumber \\
         &=& \dd s^2 + \dd X^\mu \dd X_\mu + {1\oo {2!}}
        \dd X^{\mu_1 \mu_2} \dd X_{\mu_1 \mu_2} + ... + {1\oo {n!}}
        \dd X^{\mu_1 ... \mu_n} \dd X_{\mu_1 ... \mu_n}
\lbl{19}
\ear
If $\gamma_\mu$ are taken to be dimensionless so that $X^\mu,~X^{\mu \nu}$,
etc., have respectively  dimensions of $length$, $length^2$, etc., then
a suitable power of a length parameter has to be included in every term of
eq. (\ref{19}). One natural choice is to take the length parameter equal
to the Planck scale $L_P$. For simplicity reasons we may then use the system
of units in which $L_P =1$.

In 4-dimensional spacetime the vector (\ref{12}) and its square (\ref{19})
can
be written as
\be
        \dd X = \dd s + \dd x^\mu \gamma_\mu + {1\oo 2} \dd x^{\mu \nu}
        \gamma_\mu \wedge \gamma_\nu + \dd {\tilde x}^\mu \, I \gamma_\mu
        + \dd {\tilde s} I
\lbl{20}
\ee
\be
       |\dd X |^2 = \dd s^2 + \dd x^\mu \dd x_\mu + {1\oo 2} \dd x^{\mu \nu}
       \dd x_{\mu \nu} - \dd {\tilde x}^\mu \dd {\tilde x}_\mu - \dd {\tilde
s}^2
\lbl{21}
\ee
where we now use the lower case symbols for coordinates.
The minus sign in the last two terms of the above quadratic form occurs
because in 4-dimensional spacetime with signature $(+ - - - )$
we have $I^2 = (\gamma_0 \gamma_1 \gamma_2 \gamma_3)
(\gamma_0 \gamma_1 \gamma_2 \gamma_3) = -1$, and $I^\dagger I =
(\gamma_3 \gamma_2 \gamma_1 \gamma_0)(\gamma_0 \gamma_1 \gamma_2 \gamma_3) =
-1$.

In eq.(\ref{21}) the line element $\dd x^\mu \dd x_\mu$ of the ordinary
special or general relativity is replaced by the line element
in Clifford space. A ``square root" of such a generalized line element is
$\dd X$ of eq.(\ref{20}). The latter object is a {\it polyvector},
a differential of the coordinate polyvector field
\be
         X = s + x^\mu \gamma_\mu + {1\oo 2} x^{\mu \nu} \gamma_\mu \wedge
         \gamma_\nu + {\tilde x}^\mu I \gamma_\mu + {\tilde s} I
\lbl{22}
\ee
whose square is
\be
          |X|^2 = s^2 + x^\mu x_\mu + {1\oo 2} x^{\mu \nu} x_{\mu \nu} -
          {\tilde x}^\mu {\tilde x}_\mu - {\tilde s}^2
\lbl{23}
\ee
The polyvector $X$ contains not only the vector part $x^\mu \gamma_\mu$,
but also a {\it scalar part} $s$, {\it tensor part} $x^{\mu \nu}
\gamma_\mu \wedge \gamma_\nu$, {\it pseudovector part} ${\tilde x}^\mu \,
I \gamma_\mu$ and pseudoscalar part ${\tilde s} I$. Similarly for the
differential $\dd X$.

When calculating the quadratic forms $|X|^2$ and $|\dd X|^2$ one obtains in
4-dimensional spacetime with pseudo euclidean signature $(+ - - -)$ the
minus sign in front of the squares of the pseudovector and pseudoscalar
terms. This is so, because in such a case the pseudoscalar unit square
in flat spacetime is $I^2 = I^\dagger I = -1$. In 4-dimensions
$I^\dagger = I$ regardless of the signature.

Instead of Lorentz transformations---pseudo rotations in spacetime---which
preserve $x^\mu x_\mu$ and $\dd x^\mu \dd x_\mu$ we have now more general
rotations---rotations in $C$-space---which preserve $|X|^2$ and $|\dd X|^2$.

\section{$C$-space and conformal transformations}

From (\ref{21}) and (\ref{23}) we see that a subgroup of the Clifford Group,
or rotations in
$C$-space is the group SO(4,2). The transformations of the latter group
rotate $x^\mu$, $s$, ${\tilde s}$, but leave $x^{\mu \nu}$ and ${\tilde
x}^\mu$ unchanged. Although according to our assumption physics takes place
in full $C$-space, it is very instructive first to consider a subspace of $C$-space,
that we shall call {\it conformal space} whose isometry group is SO(4,2).

Coordinates can be given arbitrary symbols. Let us now use the symbol
$\eta^\mu$ instead of $x^\mu$, and $\eta^5$,$\eta^6$ instead of ${\tilde
s}$,
$s$. In other words, instead of $(x^\mu ,{\tilde s},s)$ we write
$(\eta^\mu , \eta^5, \eta^6)
\equiv \eta^a$, $\mu=0,1,2,3$, $a = 0,1,2,3,5,6$. The quadratic form reads
\be
      \eta^a \eta_a = g_{ab} \eta^a \eta^b
\lbl{24}
\ee
with
\be
         g_{ab} = {\rm diag} (1,-1,-1,-1,-1,1)
\lbl{25}
\ee
being the diagonal metric of the flat 6-dimensional space, a subspace of
$C$-space, parametrized by coordinates $\eta^a$. The transformations which
preserve the quadratic form (\ref{24}) belong to the group SO(4,2). It
is well known \ci{Kastrup, Barut} that the latter group, when taken on
the cone
\be
       \eta^a \eta_a = 0
\lbl{26}
\ee
is identical to the 15-parameter group of conformal transformations in
4-dimensional spacetime \ci{Conformal}.

Let us consider first the rotations of $\eta^5$ and $\eta^6$ which leave
coordinates $\eta^\mu$ unchanged. The transformations that leave
$-(\eta^5)^2 + (\eta^6)^2$ invariant are
\bear
          && \eta'^5 = \eta^5 \, {\rm ch} \, \alpha + \eta^6 \, {\rm sh} \,
\alpha
          \nonumber \\
          && \eta'^6 = \eta^5 \, {\rm sh} \, \alpha + \eta^6 \, {\rm ch} \,
\alpha
\lbl{27}
\ear
where $\alpha$ is a parameter of such pseudo rotations.

Instead of the coordinates $\eta^5$, $\eta^6$ we can introduce \ci{Kastrup,
Barut}
new coordinates
$\kappa$, $\lambda$ according to
\bear
        && \kappa = \eta^5 - \eta^6 \lbl{28} \\
        && \lambda =\eta^5 + \eta^6 \lbl{29}
\ear
In the new coordinates the quadratic form (\ref{24}) reads
\be
        \eta^a \eta_a = \eta^\mu \eta_\mu - (\eta^5)^2 - (\eta^6)^2 =
        \eta^\mu \eta_\mu - \kappa \lambda
\lbl{30}
\ee
The transformation (\ref{27}) becomes
\be
         \kappa' = \rho^{-1} \kappa   \lbl{31} \ee
\be
        \lambda' = \rho \lambda     \lbl{32} \ee
where $\rho = e^\alpha$. This is just a dilation of $\kappa$
and the inverse dilation of $\lambda$.

Let us now introduce new coordinates $x^\mu$ according $x^\mu$
to\footnote{These new coordinates $x^\mu$ should not be confused with
coordinate $x^\mu$ used in Sec.2.}
\be
         \eta^\mu = \kappa x^\mu
\lbl{33}
\ee
Under the transformation (\ref{33}) we have
\be
        \eta'^\mu = \eta^\mu
\lbl{33a}
\ee
but
\be
         x'^\mu = \rho x^\mu
\lbl{34}
\ee
The latter transformation is {\it dilatation} of coordinates $x^\mu$.

Considering now a line element
\be
       \dd \eta^a \dd \eta_a = \dd \eta^\mu \dd \eta_\mu - \dd \kappa \dd
\lambda
\lbl{33b}
\ee
we find that {\it on the cone} $\eta^a \eta_a =0$ it is
\be
           \dd \eta^a \dd \eta_a = \kappa^2 \, \dd x^\mu \dd x_\mu
\lbl{35}
\ee
even if $\kappa$ is not constant.
Under the transformation (\ref{31}) we have
\be
           \dd \eta'^a \dd \eta'_a = \dd \eta^a \dd \eta_a
\lbl{36}
\ee
\be
             \dd x'^\mu \dd x'_\mu = \rho^2 \, \dd x^\mu \dd x_\mu
\lbl{37}
\ee
The last relation is a {\it dilatation} of the 4-dimensional line element
related
to coordinates $x^\mu$. In a similar way also other transformations
of the group SO(4,2) that preserve
(\ref{26}) and (\ref{36}) we can rewrite in terms of of the coordinates
$x^\mu$.
So we obtain---besides dilations---translations, Lorentz transformations,
and special conformal transformations; altogether they are called {\it
conformal
transformations}.  This is a well known old
observation \ci{Kastrup, Barut} and we shall not discuss it further.
What we wanted to point out here is that conformal group
SO(4,2) is a subgroup of the Clifford group.

\section{On the physical interpretation of the conformal group SO(4,2)}

In order to understand the physical meaning of the transformations
(\ref{33})
from the coordinates $\eta^\mu$ to the coordinates $x^\mu$ let us consider
the following transformation in 6-dimensional space $V_6 \,$:
\bear
             && x^\mu = \kappa^{-1} \eta^\mu \nonumber \\
             && \alpha = - \kappa^{-1} \nonumber \\
             && \Lambda = \lambda - \kappa^{-1} \eta^\mu \eta_\mu
\lbl{38}
\ear
This is a transformation from the coordinates $\eta^a = (\eta^\mu,
\kappa,\lambda)$
to the new coordinates $x^a = (x^\mu , \alpha, \Lambda)$. No extra condition
on coordinates, such as (\ref{26}), is assumed now. If we calculate the line
element in the coordinates $\eta^a$ and $x^a$, respectively, we find the
the following relation \ci{Pavsic3}
\be
       \dd \eta^\mu \dd \eta^\nu \, g_{\mu \nu} - \dd \kappa \, \dd \lambda 
=
        \alpha^{-2} (\dd x^\mu \dd x^\nu \, g_{\mu \nu} - \dd \alpha
        \dd \Lambda )
\lbl{39}
\ee

We can interpret a transformation of coordinates passively or actively.
Geometric calculus clarifies significantly the meaning of passive and
active transformations. Under a {\it passive transformation} a vector
remains the same, but its components and basis vector change. For
a vector $\dd \eta = \dd \eta^a \gamma_a$ we have
\be
           \dd \eta' = \dd \eta'^a \gamma'_a = \dd \eta^a \gamma_a = \dd 
\eta
\lbl{40}
\ee
with
\be
                  \dd \eta'^a  = {{\p \eta'^a}\oo {\p \eta^b}} \, \dd  
\eta^b
\lbl{41}
\ee
and
\be
        \gamma'_a = {{\p \eta^b}\oo {\p \eta'^a}} \, \gamma_b
\lbl{42}
\ee
Since the vector is invariant, so it is its square:
\be
       \dd \eta'^2 = \dd \eta'^a \gamma'_a \, \dd \eta'^b \gamma'_b =  \dd
\eta'^a
      \dd \eta'^b g'_{ab}  =  \dd \eta^a  \dd \eta^b g_{ab}
\lbl{42a}
\ee
From (\ref{42}) we read that the well known relation between  new and
old coordinates:
\be
        g'_{ab} = {{\p \eta^c }\oo {\p \eta'^a}} \, {{\p \eta^d}\oo {\p
\eta'^b}}
       \, g_{cd}
\lbl{43}
\ee

Under an {\it active transformation} a vector changes. This means that in a
fixed basis the components of a vector change:
\be
          \dd \eta' = \dd \eta'^a \gamma_a
\lbl{43a}
\ee
with
\be
          \dd \eta'^a  = {{\p \eta'^a}\oo {\p \eta^b}} \, \dd  \eta^b
\lbl{44}
\ee
The transformed vector $\dd \eta'$ is different from the original vector
$\dd \eta = \dd \eta^a \gamma_a$. For the square we find
\be
       \dd \eta'^2 = \dd \eta'^a \dd \eta'^b g_{ab} = {{\p \eta'^a} \oo {\p
\eta^c}}
         {{\p \eta'^b} \oo {\p \eta^d}} \, \dd \eta^c \dd \eta^d g_{ab}
\lbl{45}
\ee
i.e., the transformed line element $\dd \eta'^2$ is different from the
original
line element.

Returning now to the coordinate transformation (\ref{38}) with the
identification
$\eta'^a = x^a$, we can interpret eq. (\ref{39}) passively or actively.

In  the {\it passive interpretation} the metric tensor and the components
$\dd \eta^a$ change under a transformation, so that in our particular
case the relation (\ref{42a}) becomes
\be
        \dd x^a \, \dd x^b \, g'_{ab} = \alpha^{-2}(\dd x^\mu \dd x^\nu \,
        g_{\mu \nu} - \dd \alpha \, \dd \Lambda ) = \dd \eta^a \dd \eta^b
        g_{ab} = \dd \eta^\mu \dd \eta^\nu g_{\mu \nu} - \dd \kappa \, \dd
        \lambda
\lbl{46}
\ee
with
\be
            g'_{ab} = \alpha^{-2} \pmatrix{g_{\mu \nu} & 0 & 0 \cr
                                   0 &       0 & -{1\oo 2} \cr
                                   0 &   -{1\oo 2} & 0 \cr }
    \; \quad , \qquad g_{ab} = \pmatrix{g_{\mu \nu} & 0 & 0 \cr
                                   0 &       0 & -{1\oo 2} \cr
                                   0 &   -{1\oo 2} & 0 \cr }
\lbl{47}
\ee
In the above equation the same infinitesimal distance squared is expressed
in
two different coordinates $\eta^a$ or $x^a$.

In {\it active interpretation}, only $\dd \eta^a$ change, whilst the metric
remains the same, so that the transformed element is
\be
         \dd x^a \, \dd x^b \, g_{ab} = \dd x^\mu \dd x^\nu \,
        g_{\mu \nu} - \dd \alpha \, \dd \Lambda = \kappa^{-2} \,
        \dd \eta^a \dd \eta^b g_{ab} = \kappa^{-2}
        (\dd \eta^\mu \dd \eta^\nu g_{\mu \nu} - \dd \kappa \, \dd
        \lambda )
\lbl{48}
\ee
The transformed line lelement $\dd x^a \dd x_a$ is physically different
from the original line element $\dd \eta^a \dd \eta_a$ by a factor
$\alpha^2 = \kappa^{-2}$

A rotation (\ref{27}) in the plane $(\eta^5, \eta^6)$ (i.e.,
the transformation (\ref{31}),(\ref{32}) of $(\kappa , \lambda))$ manifests
in the new coordinates $x^a$ as a {\it dilatation} of the line element
$\dd x^a \dd x_a = \kappa^{-2} \, \dd \eta^a \eta_a$:
\be
           \dd x'^a \dd x'_a  = \rho^2 \dd x^a \dd x_a
\lbl{50}
\ee

All this is true in the full space $V_6$. On the cone $\eta^a \eta_a = 0$ we
have $\Lambda = \lambda - \kappa \eta^\mu \eta_\mu = 0$, $\dd \Lambda = 0$
so that $\dd x^a \dd x_a = \dd x^\mu \dd x_\mu$ and we reproduce the
relations (\ref{37}) which is a dilatation of the 4-dimensional line
element.
It can be interpreted either passively or actively. In general, the
pseudo rotations in $V_n$, that is, the transformations of the 15-parameter
group SO(4,2) when expressed in terms of coordinates $x^a$, assume on the
cone $\eta^a \eta_a = 0$ the form of the ordinary conformal transformations.
They all can be given the active interpretation \ci{Pavsic2, Pavsic3}.

\section{Conclusion}

We started from the new paradigm that physical phenomena actually occur
not in spacetime, but in a larger space, the so called {\it Clifford space}
or $C$-space which is a manifold associated with the Clifford algebra
generated by the basis vectors $\gamma_\mu$ of spacetime. An arbitrary
element
of Cliffod algebra can be expanded in terms of the objects $E_A$ , $A =
1,2,...,
2^D$,
which include, when $D=4$, the scalar unit $\bf 1$,
vectors $\gamma_\mu$, bivectors $\gamma_\mu \wedge \gamma_\nu$,
pseudovectors $I \gamma_\mu$ and the pseudoscalar unit $I \equiv \gamma_5$.
$C$-space contains 6-dimensional subspace $V_6$ spanned\footnote{
It is an old observation \ci{Mirman} that the generators
$L_{ab}$ of SO(4,2) can be realized in
terms of ${\bf 1}$, $\gamma_\mu$, and $\gamma_5$. Lorentz generators are
$M_{\mu \nu} = -{i\oo 4} [\gamma_\mu , \gamma_\nu]$, dilatations are
generated
by $D=L_{65} =-{1\oo 2} \gamma_5$, translations by $P_\mu = L_{5 \mu} +
L_{6 \mu} = {1\oo 2} \gamma_\mu (1- i \gamma_5)$ and the special
conformal transformations by $L_{5 \mu} - L_{6 \mu} =
{1\oo 2} \gamma_\mu (1+ i \gamma_5)$. This essentially means
that the generators are $L_{ab} = -{i\oo 4} [e_a,e_b]$ with $e_a =
(\gamma_\mu,
\gamma_5, {\bf 1})$, where care must be taken to replace commutators
$[{\bf 1}, \gamma_5]$ and $[{\bf 1}, \gamma_\mu]$ with $2 \gamma_5$ and
$2 \gamma_\mu$}
by ${\bf 1}$, $\gamma_\mu$,
and $\gamma_5$. The metric of $V_6$ has the signature $(+----+)$. It is
well known that the rotations
in $V_6$, when taken on the conformal cone $\eta^a \eta_a = 0$,
are isomorphic to the non linear transformations of the conformal group
in spacetime. Thus we have found out that $C$-space contains ---as
a subspace---
the 6-dimensional space $V_6$ in which the conformal group acts linearly.
From the physical point of view this is an important and, as far as we
know, a novel finding, although it might look mathematically trivial.
So far it has not been clear what could be a physical interpretation
of the 6 dimensional conformal space. Now we see that it is just
a subspace of Clifford space.

We take $C$-space seriously as an arena in which physics takes place.
The theory is a very natural, although not trivial, extension of the
special relativity in spacetime. In special relativity the transformations
that preserve the quadratic form are given an {\it active interpretation}:
they relate the objects or the systems of reference in {\it relative
translational motion}. Analogously also the transformations that preserve
the quadratic form (\ref{21}) or (\ref{23}) in $C$-space should
be given an active interpretation. We have found that among such
transformations (rotations in $C$-space) there exist the
transformations of the group SO(4,2). Those transformations also should
be given an active interpretation as the transformations that relate
different physical objects or reference frames. Since in the ordinary
relativity we do not impose any constraint on the coordinates of a freely
moving object so we should not impose any constraint in $C$-space, or in
the subspace $V_6$. However, by using the projective coordinate
transformation
(\ref{38}), without any constraint such as  $\eta^a \eta_a = 0$, we arrived
at the relation (\ref{48})  for the line elements. If in the coordinates
$\eta^a$ the line element is constant, then in the coordinates $x^a$ the
line element is changing by a scale factor $\kappa$ which, in general,
depends on the evolution parameter $\tau$. The line element does not
necessarily
relate the events along a particle's worldline. We may consider the line
element between two infinitesimally separated events within an extended
object where both have the same coordinate label $\Lambda$
so that $\dd \Lambda =0$. Then the 6-dimensional
line element  $\dd x^\mu \dd x^\nu \, g_{\mu \nu} - \dd \alpha \, \dd
\Lambda$
becomes the 4-dimensional line element $\dd x^\mu \dd x^\nu \, g_{\mu \nu}$
and, because of (\ref{48}) it changes with $\tau$ when $\kappa$ does
change. This means that the object  changes its size, it is moving
dilatationally \cite{Pavsic2,Pavsic3}.
We have thus arrived at a very far reaching
observation that the relativity in $C$-space implies {\it scale changes
of physical objects as a result of free motion, without presence of
any forces or such fields as assumed in Weyl theory}. This was advocated 
long
time ago \cite{Pavsic2,Pavsic3}, but without recurse to $C$-space.

An immediate project would be to construct the gauge theory associated with 
the Clifford algebra of spacetime. This is currently under
investigation and has been called the C-space generalization of Maxwell's 
Electromagnetism which describes the dynamics and couplings of extended 
objects to antisymmetric tensor fields of arbitrary rank.
\cite{Antonio}.

\section{Acknowledgements}

We are indebted to Prof. David Finkelstein for his insighful comments and 
reading of the manuscript.

\end{document}